\documentclass[12pt]{article}
\input epsf

\begin{document}

\begin{center}
{\Large\bf Storage in Computational Geometry}

\medskip

{\it Yijie Han$^1$} and {\it Sanjeev Saxena$^2$}

\medskip

$^1$School of Science and Engineering

University of Missouri at Kansas City

Kansas City, Missouri 64110

hanyij@umkc.edu

\medskip

$^2$Computer Science and Engineering

Indian Institute of Technology

Kanpur-208 016, India

ssax@cse.iitk.ac.i

\end{center}

\begin{abstract}
We show that $n$ real numbers can be stored in 
a constant number of real numbers 
such that each original real number can be fetched in $O(\log n)$ time.

Although our result has implications for many computational geometry problems,
we show here, combined with Han's $O(n\sqrt{\log n})$ time real number sorting algorithm \cite{HanRealNumber}, we can
improve the complexity of Kirkpatrick's point location algorithm \cite{Kirkpatrick} to
$O(n\sqrt{\log n})$ preprocessing time, a constant number of real numbers 
for storage and $O(\log n)$ point
location time. Kirkpatrick's algorithm uses
$O(n\log n)$ preprocessing
time, $O(n)$ storage and $O(\log n)$ point location time. The complexity results in Kirkpatrick's algorithm was the previous best result. Although Lipton and
Tarjan's algorithm
\cite{Lipton} predates Kirkpatrick's algorithm and has the same complexity,
Kirkpatrick's algorithm is simpler and has a better structure.

This paper can be viewed as a companion paper of paper \cite{HanRealNumber}.\\

\end{abstract}

\baselineskip=4mm

\section{Introduction}

In the past it is usually thought that multiple real numbers cannot be packed
into fewer real numbers with small access time. Thus many computational geometry
algorithms working with $O(n)$ size problems use $O(n)$ storage to store 
$O(n)$ real numbers. 

We show, on the contrary, $n$ real numbers can be packed into a constant number
of real numbers 
such that each original real number can be accessed in $O(\log n)$
time. This result enables us to reduce the storage requirement for
many computational geometry problems to a constant number of real numbers. 

What we have reached is to store 6 real numbers into 5 real numbers and 2 rational numbers.
The reader can first try by himself/herself to see if he/she can store $c$ real numbers into $c-1$ real numbers
for an integer $c>1$ such that each original real number can be fetched in constant time. 
This solution is not trivial and we worked on it for some time to resolve it.

In particular, we show, combined with Han's $O(n\sqrt{\log n})$ time
real number sorting
algorithm \cite{HanRealNumber}, 
we can improve the complexity of Kirkpatrick's point location
algorithm \cite{Kirkpatrick} to
$O(n\sqrt{\log n})$ preprocess time, a constant number of real numbers 
for storage
and $O(\log n)$ point location time. Kirkpatrick's algorithm \cite{Kirkpatrick} has
$O(n\log n)$ preprocessing time, $O(n)$ storage and $O(\log n)$ point location
time. 

Although Lipton and Tarjan's algorithm \cite{Lipton} for point location
predates Kirkpatrick's algorithm and has the same complexity as Kirkpatrick's
algorithm,
we choose to apply our result to Kirkpatrick's
algorithm \cite{Kirkpatrick} because it is more applicable. 

\section{Store 6 Real Numbers into 5 Real Numbers and 2 Rational Number}
We use 6 lines $L_1, L_2, L_3, L_4, L_5, L_6$ on the plane with these lines' slopes being these 6 real numbers.
We let $L_1, L_2, L_3$ intersect at a point $A$ and let $L_4, L_5, L_6$ intersect at another point $B$ as shown in Fig. 1.

\begin{figure}
\epsfxsize=4in
\centerline{\epsffile{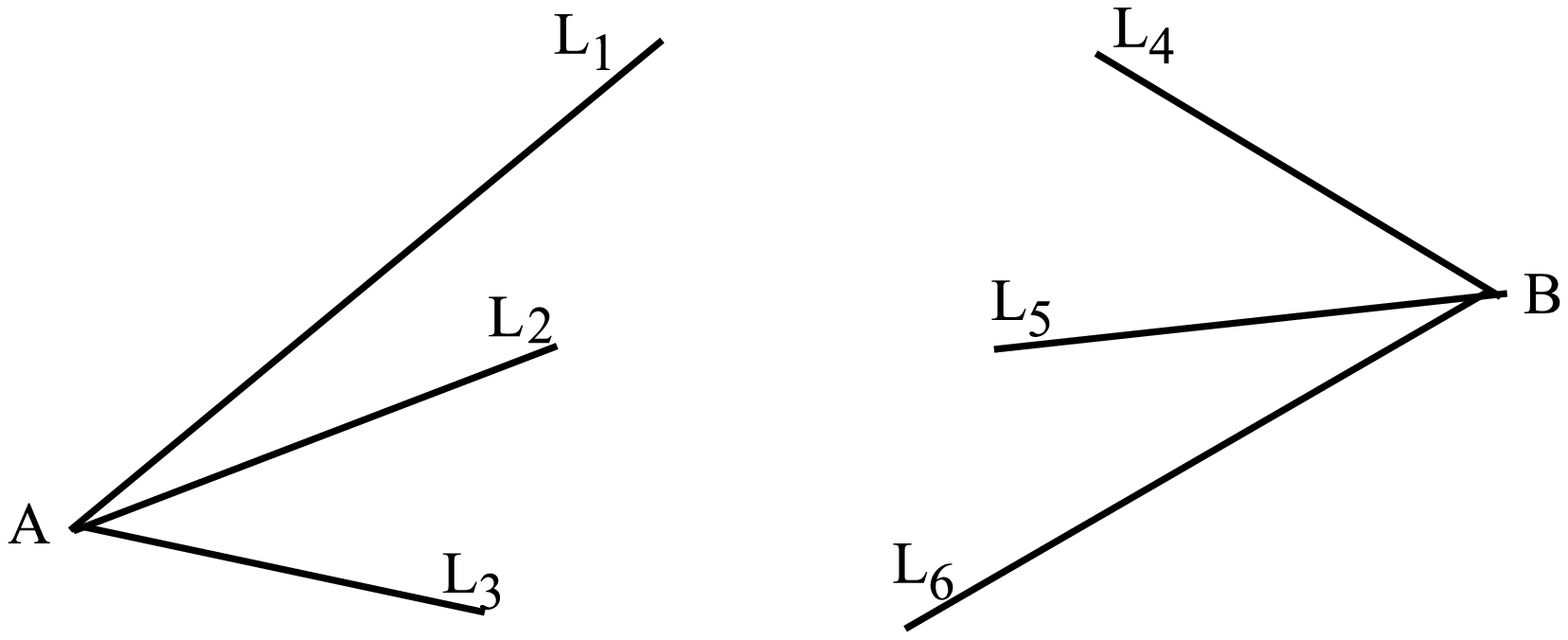}}
\centerline{Fig. 1.}
\end{figure}

We draw lines $L^\prime$ and $L^{\prime \prime}$ with the same slope and $L^\prime$ intersects $L_1, L_2, L_3$ at $a, b, c$ and 
$L^{\prime \prime}$ intersects
$L_4, L_5, L_6$ at $d, e, f$ as shown in Fig. 2. If we tilt $L^\prime$ and $L^{\prime \prime}$ simultaneously but keep their slopes equal then $length(a, b)/length(b, c)$ increases and $length(d, e)/length(e, f)$ decreases ($length(a, b)/length(b, c)$ decreases and $length(d, e)/length(e, f)$ increases). To see this take a look
at Fig. 3. In Fig. 3 we let line L rotate around point $c$. We have
$sinC/length(a, c)=sinA/length(b, c)$ and $sinD/length(a, c)=sinB/length(c, d)$.
Thus $length(b, c)/length(c, d)=(sinA*length(a, c)/sinC)/(sinB*length(a, c)/sinD)=sinAsinD/sinBsinC=(sinA/sinB)(sin(A+B+C)/sinC)=(sinA/sinB)(sin(A+B)cosC/sinc+cos(A+B))$. When we rotate line $L$ around point $c$ counter-clockwise from 0 degree slope to 180 degree, slope $C$ changes from 0 degree to 180 degree and $A$ and $B$ are constants. Thus $length(b, c)/length(c, d)$ is strictly decreasing.

\begin{figure}
\epsfxsize=4in
\centerline{\epsffile{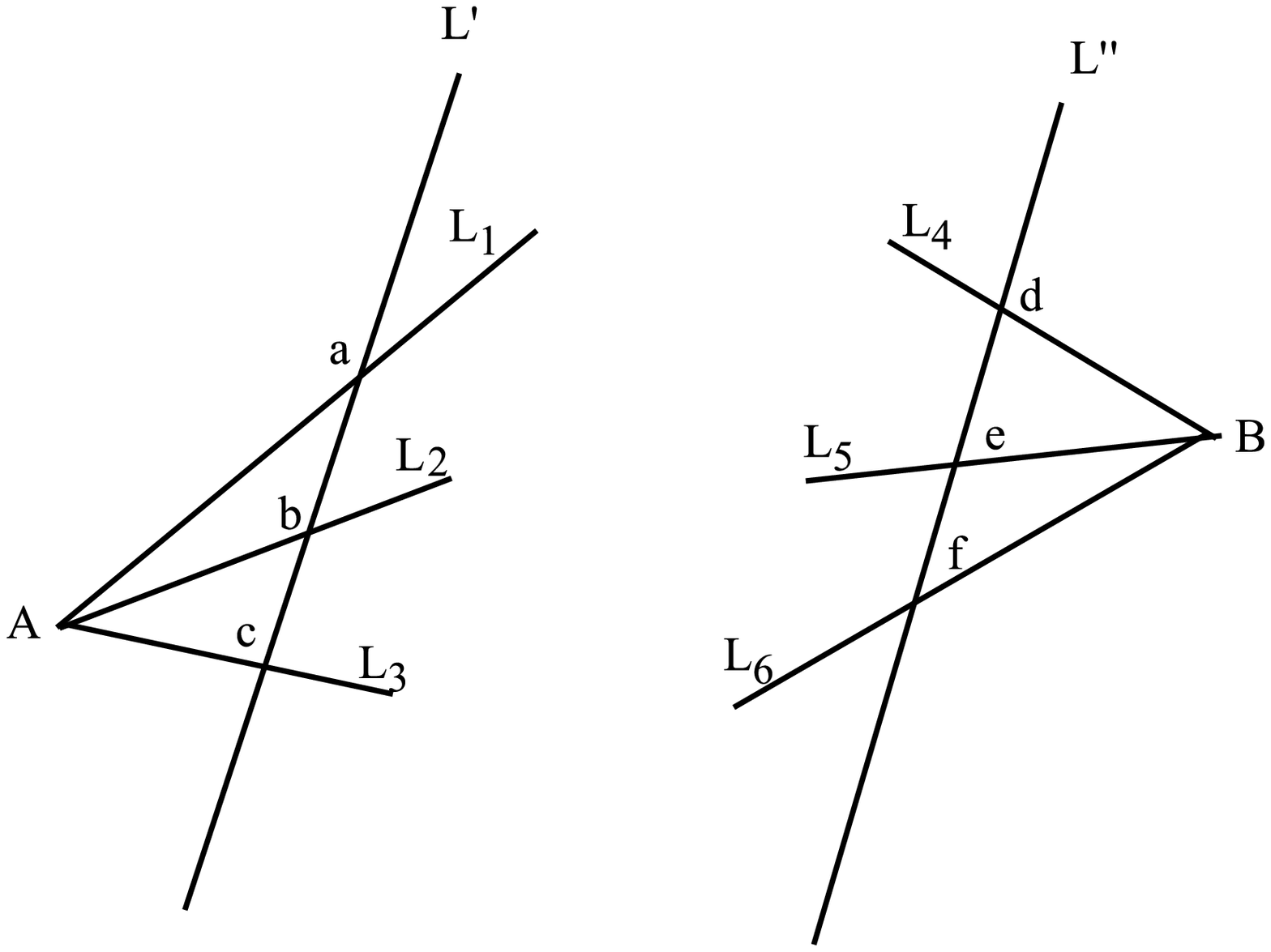}}
\centerline{Fig. 2.}
\end{figure}

\begin{figure}
\epsfxsize=2.5in
\centerline{\epsffile{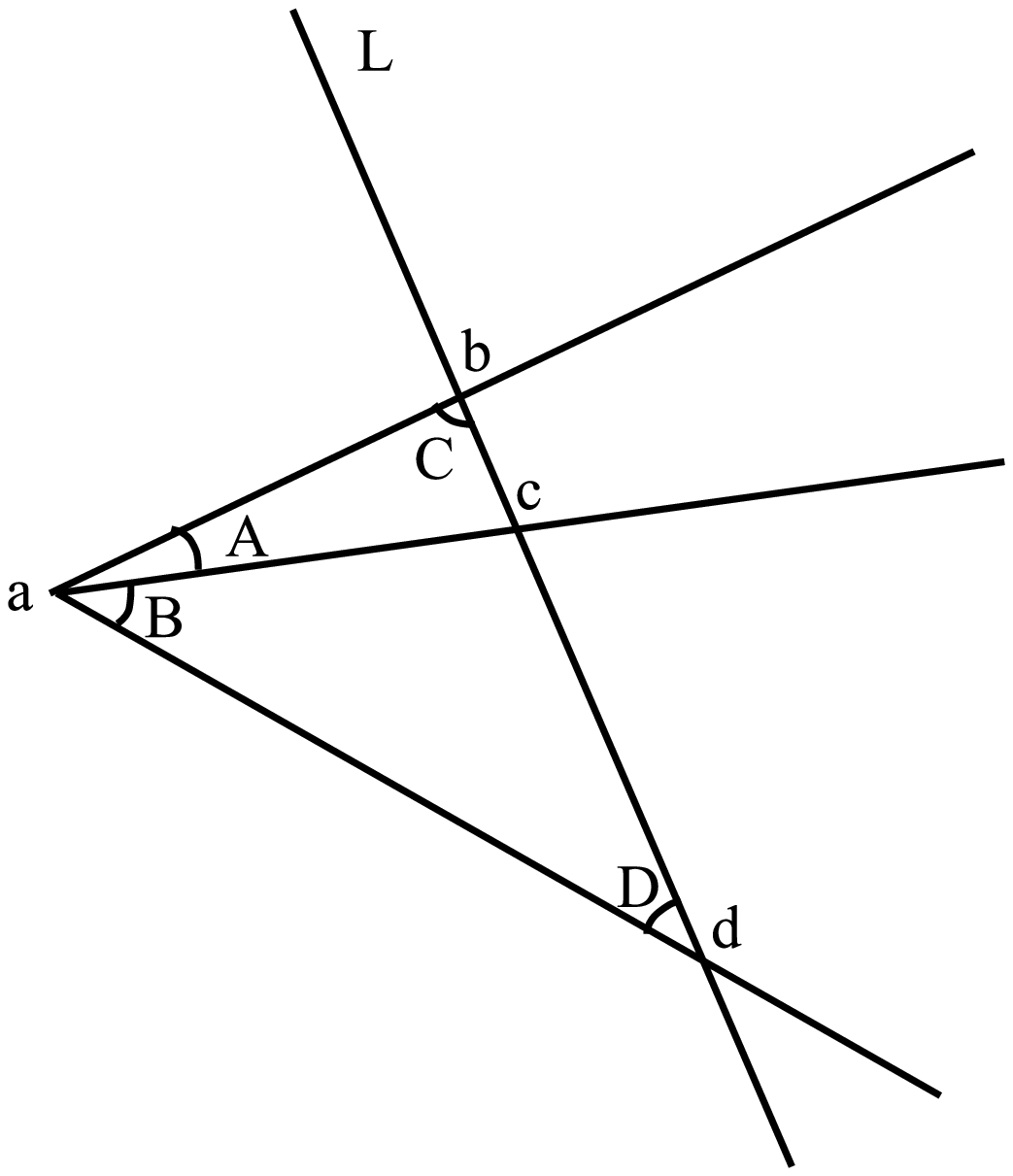}}
\centerline{Fig. 3.}
\end{figure}

Thus there is a slope $s$ for $L^\prime$ and $L^{\prime \prime}$ such that $length(a, b)/length(b, c)=length(d, e)/length(e, f)$. We will fix this slope $s$ and combine $L_1, L_2, L_3$ and $L_4, L_5, L_6$ together as shown in Fig. 4. In Fig. 4. we replace $L^\prime$ and $L^{\prime \prime}$ by line $L$. Point $a$ and $d$ will be combined into $a$, $b$ and $e$ will be combined into $b$ and $c$ and $f$ will be
combined into $c$.

\begin{figure}
\epsfxsize=4in
\centerline{\epsffile{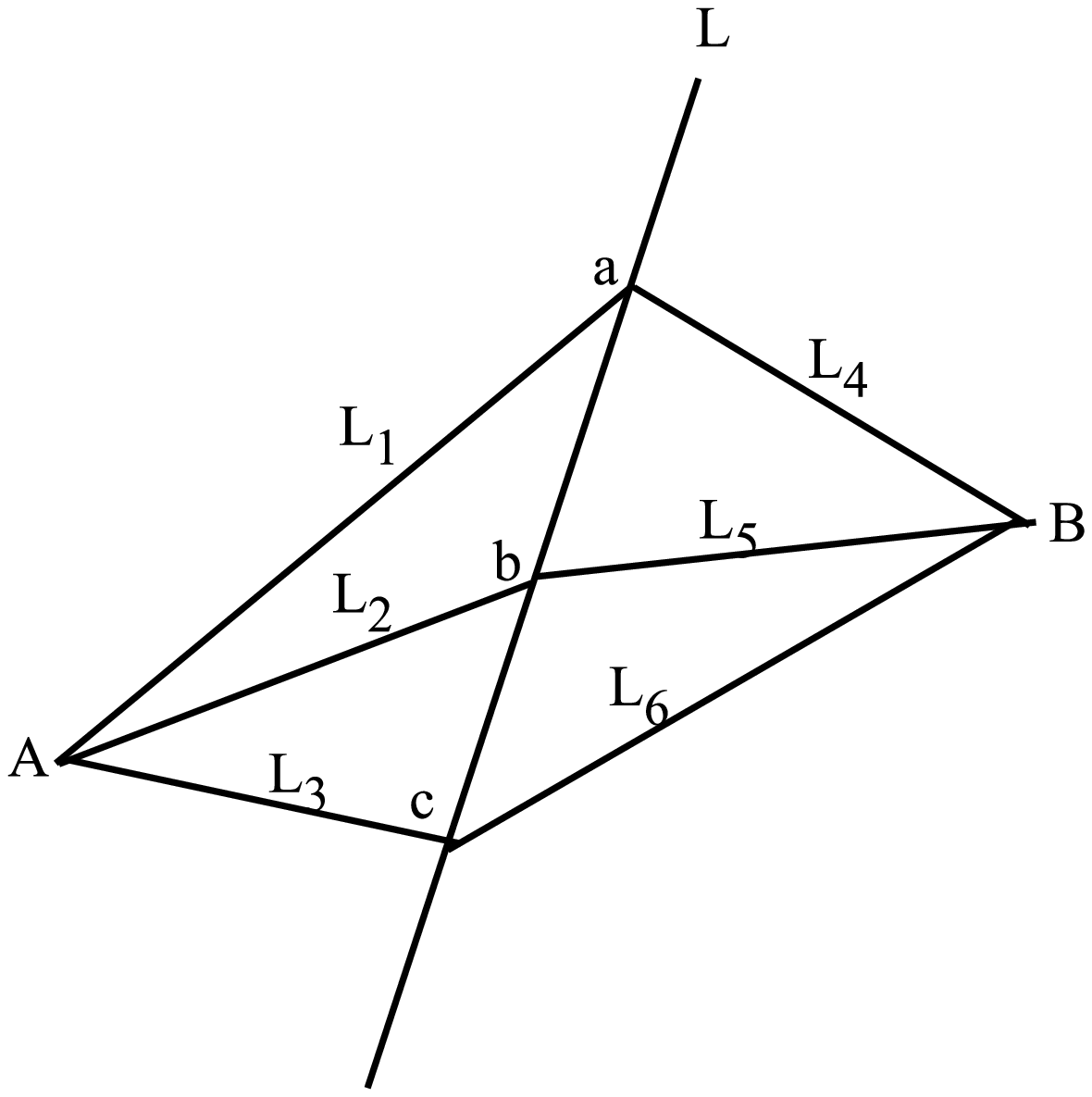}}
\centerline{Fig. 4.}
\end{figure}

We now place point $A$ at $(0, 0)$. Because we only need the slopes of lines and therefore Fig. 4 can be scaled to larger or smaller without affecting the slopes of lines. If we fix point $A$ at $(0, 0)$ and continuously scale Fig. 4 then the trace of point $B$ is a continuous line $C$ and any point on $C$ can be used for the position of $B$. Because $C$ is a continuous line therefore there is a point $(i, j)$ on $C$ with one of $i$ and $j$ being a rational number. We will fix
this $(i, j)$ as the position of $B$. This is shown as in Fig. 5.

\begin{figure}
\epsfxsize=4in
\centerline{\epsffile{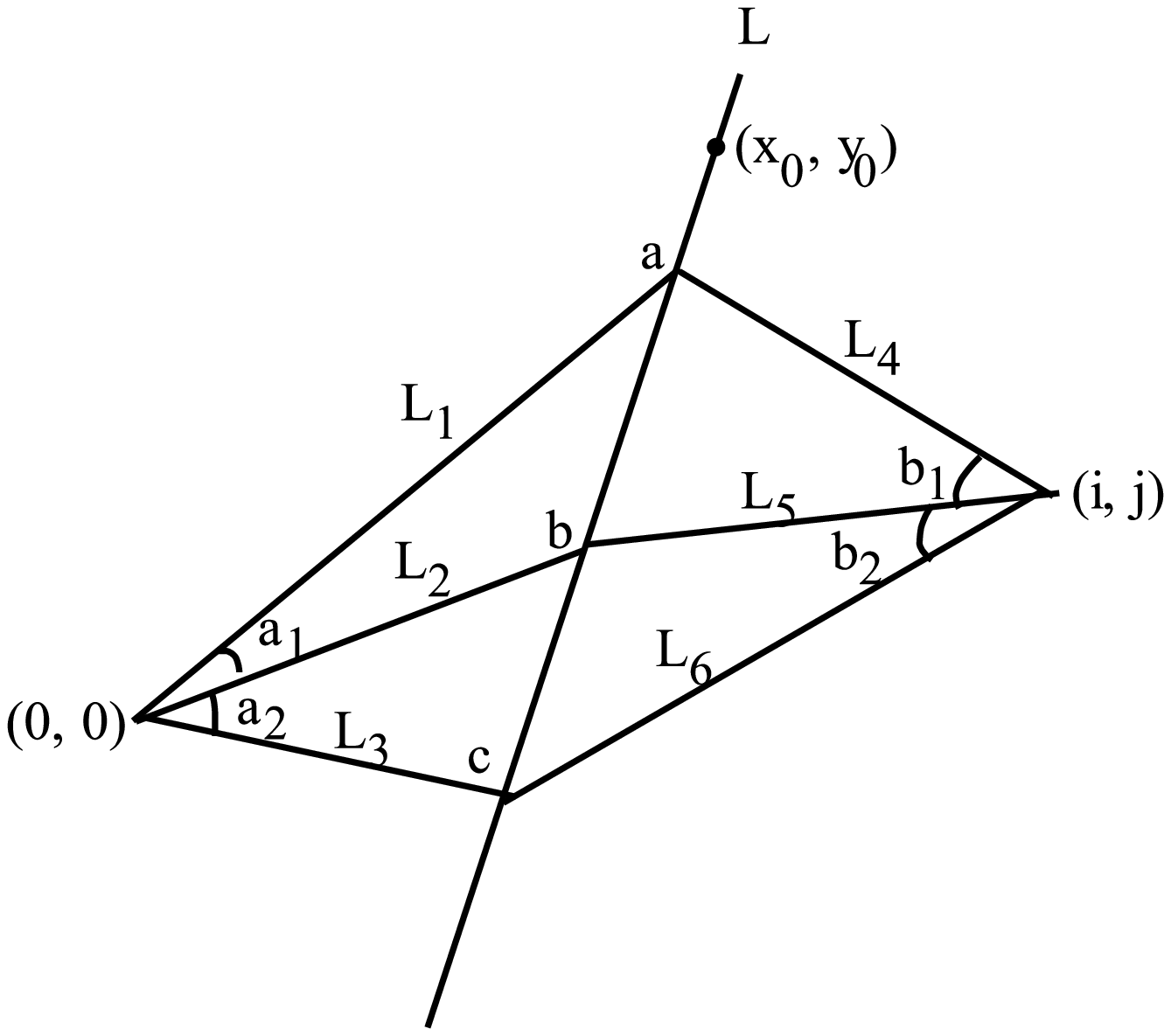}}
\centerline{Fig. 5.}
\end{figure}

We will store 5 real numbers (see Fig. 5.): angle $b_1$, angle $b_2$, $length(a, b)/length(b, c)$, one of $i$ and $j$ that is a real number (other one is a rational number) and one of $(x_0, y_0)$ that is a real number, where $(x_0, y_0)$ is any point on $L$ and
one of $x_0$ and $y_0$ is a rational number. Here we used 5 real numbers and 2 rational numbers. 

We show from these stored 5 real numbers and 2 rational number we can recover the 6 real numbers that are the slopes of lines $L_1$ to $L_6$.

Because we know $b_1$, $b_2$ and $length(a, b)/length(b, c)$ therefore the angle between $L$ and $L_4$, the angle between $L$ and $L_5$, the angle between $L$
and $L_6$ are fixed. Because $L_4, L_5, L_6$ intersect at $(i, j)$ and $L$ passes $(x_0, y_0)$ and thus lines $L, L_4, L_5, L_6$ are fixed. Thus points $a, b, c$
are fixed. Now lines $L_1, L_2, L_3$ can be obtained by connecting $(0, 0)$
with $a$, connecting $(0, 0)$ with $b$, connecting $(0, 0)$ with $c$.
   
\section{Accessing Real Numbers}
We can pack 18 real numbers into 15 real numbers and 6 rational numbers.
These 6 rational numbers $r_1=a_1/b_1, r_2=a_2/b_2, r_3=a_3/b_3, r_4=a_4/b_4, r_5=a_5/b_5, r_6=a_6/b_6$, where $a_1, b_1, a_2, b_2, a_3, b_3, a_4, b_4, a_5, b_5, a_6, b_6$ are integers in 2's complement. Let $k$ be the smallest integer such that
$2^k \geq {\rm max} \{2a_1, 2b_1, 2a_2, 2b_2, 2a_3, 2b_3, 2a_4, 2b_4, 2a_5, 2b_5, 2a_6, 2b_6\}$ (here we view $a_i, b_i$, $1 \leq i \leq 6$, as unsigned integers). We store $k$ as one number and store $a_1*2^{11k}+b_1*2^{10k}+a_2*2^{9k}+b_2*2^{8k}+a_3*2^{7k}+b_3*2^{6k}+a_4*2^{5k}+b_4*2^{4k}+a_5*2^{3k}
+b_5*2^{2k}+a_6*2^k+b_6$ as another number. Thus we stored 18 real numbers in 17 real numbers.

The packing of $n$ real numbers to a constant number of real numbers
forms a tree like hierarchy with $\lceil \log_{18/17} n\rceil$ level with
each level packing every 18 real numbers to 17 real numbers.

Thus to access a real number at a leaf of the tree we follow the path from
the root of the tree to the leaf of the tree with constant decomposition
time to decompose 17 real numbers to 18 real numbers.\\

\noindent
{\bf Theorem 1:} $n$ real numbers can be packed into a constant number of
real numbers with $O(\log n)$ access time for each original real number.
$\framebox{}$ 

\section{Fast Real Number Sorting Algorithm}
Comparison sorting is known to have $\Omega (n\log n)$ time lower bound \cite{MIT}. In the past points on the plane or in space are sorted by comparison
sorting as they are real numbers. Classical comparison 
sorting algorithms such as
quick sort \cite{Hoare}, merge sort \cite{Knuth} and heap sort \cite{Williams} have optimal time for comparison sorting
and they can sort points in $O(n\log n)$ time.  

However, Han presented an $O(n\sqrt{\log n})$ time real number 
sorting algorithm in
\cite{HanRealNumber}. The computation model used for this algorithm is
the same model used in computational geometry. Thus this algorithm can
be used to sort points in $O(n\sqrt{\log n})$ time.

The approach used in \cite{HanRealNumber} is to convert real numbers to
integers while preserving their order. Han showed in \cite{HanRealNumber}
that we can convert $n$ real numbers to integers while preserving
the original order of these real numbers in $O(n\sqrt{\log n})$ time
\cite{HanRealNumber}.  

After real numbers are converted into integers these $n$ integers
can be sorted in $O(n\log \log n)$ time \cite{Hanlln,HanOP}.

\section{Improve the Complexity in Kirkpatrick's Point Location Algorithm}

The bottleneck of the preprocessing time in Kirkpatrick's point location
algorithm \cite{Kirkpatrick} is sorting the $O(n)$ real numbers (point coordinates or line slopes). This was done in Kirkpatrick's algorithm using
comparison sorting. Here we use Han's $O(n\sqrt{\log n})$ time real number sorting algorithm \cite{HanRealNumber} to improve the preprocess time to
$O(n \sqrt{\log n})$.

The storage complexity for the input subdivision is $O(n)$ in Kirkpatrick's
algorithm \cite{Kirkpatrick} and we cannot improve on this.

After preprocessing Kirkpatrick stored the processed subdivision with $O(n)$
storage complexity \cite{Kirkpatrick}. Use our storage method described earlier
in this paper we can
pack the processed division into constant storage. We need to follow 
Kirkpatrick's algorithm. Kirkpatrick's algorithm forms a tree like hierarchy.
A slightly modified version of this tree has a triangle at each node $a$ of
this tree and $a$ has a constant number of children each of which is a 
triangle. The formation in Kirkpatrick's paper seems in reverse of the
tree we described here. At node $a$ the query point is located within the
triangle at $a$ and Kirkpatrick's algorithm will decide which child's (in Kirkpatrick's paper which parent's) triangle contains the query point. If we
pack real numbers following the formation of this tree, i.e. the packing of
the data at $a$ is done after packing data of $a$'s descendants, then we will
be able to extract data at $a$ and data at $a$'s children before we search downwards toward a leaf. Thus the tree can be stored in a constant number of
real numbers and support the search (point location) in $O(\log n)$ time.

We can associate each triangle in Kirkpatrick's hierarchy with 17 real numbers and 6 real numbers for the coordinate values of triangle's three points. So there are 23 real numbers associated with each triangle.
At a level of Kirkpatrick's hierarchy a triangle $A$ has a constant number of children triangles. There are 23 real numbers associated with each child triangle. We pack all the real numbers associated with children triangles into 17 real numbers and associate these 17 real numbers and the 6 real numbers of $A$'s coordinate values with $A$. \\ 

\noindent
{\bf Theorem 2:} Point location can be computed with $O(n\sqrt{\log n})$ preprocess time. After the computation the data can be stored in constant storage to facilitate $O(\log n)$ point location time. $\framebox{}$ 

\section{Related and Future Research}

In \cite{HanSairam} Chaganti and Han
were able to solve the point location problem with $O(n\sqrt{\log n})$
preprocessing time, $O(n)$ storage and constant point location time
in . The approach used there is to convert real numbers to
integers while preserving their order. Multiple integers can then be packed
into one word. The difference between the packing of integers in \cite{HanSairam}
and the packing of real numbers here is that in \cite{HanSairam} integer addition, subtraction, multiplication between one integer with $n$ integers packed in a word can be computed in constant time. Pairwise addition or multiplication between $n$ integers packed in
a word and $n$ integers packed in another word can also be computed in constant time \cite{HanSairam} (these methods
were known before us). However for the real number packing we used here we cannot do this. 

The problem with the algorithm in \cite{HanSairam} is that 
logical (bitwise) AND operation for integers is used there and is counted as a constant
time operation. That maybe not allowed in computational geometry.

We would like to investigate in the future whether real number packing can
allow the above mentioned multiple real number addition, subtraction and
multiplication to be carried in constant time. Other arithmetic and logical
operations can also be considered. If this is achievable that means $O(\log n)$ time binary search among $n$ sorted real numbers is not optimal.

Searching in constant time was presented in \cite{HanSwathi} in which 
multiple integers converted from real numbers (while preserving order) i
are packed into one word.
Again logical AND operation is used there. Although logical exclusive-or operation for integers is also used in \cite{HanSwathi} this is not required and the 
constant time searching algorithm in \cite{HanSwathi} can do without it.  

Another problem we are interested in studying is whether access time can be reduced to
$o(\log n)$ after real numbers are packed.   

\section{Conclusions}
Somewhat surprisingly we showed how to pack $n$ real numbers into constant number of real numbers. We look forward to witnessing future development along this 
line.

\end{document}